\documentclass[prb,preprint,showpacs,preprintnumbers,amsmath,amssymb]{revtex4}

\usepackage{graphicx}

\usepackage{dcolumn}

\usepackage{bm}

\newcommand{\gma}{$\rm{Ga_{1-x}Mn_{x}As}$}

\newcommand{\rxx}{$R_{\rm{xx}}$}
\newcommand{\rxy}{$R_{\rm{xy}}$}
\newcommand{\tc}{$T_{\rm{C}}$}
\newcommand{\hc}{$H_{\rm{C}}$}

\begin{document}
\title{Non-collinear Spin Valve Effect in Ferromagnetic Semiconductor Trilayers}
\author{G.\ Xiang}
\author{B. L. Sheu}
\author{M. Zhu}
\author{P. Schiffer}
\author{N.\ Samarth}\email{nsamarth@psu.edu}
\affiliation{Physics Department $\&$ Materials Research Institute, The Pennsylvania State University, University Park PA 16802}

\begin{abstract}
We report the observation of the spin valve effect in (Ga,Mn)As/p-GaAs/(Ga,Mn)As trilayer devices. Magnetoresistance measurements carried out in the current in plane geometry reveal positive magnetoresistance peaks when the two ferromagnetic layers are magnetized orthogonal to each other. Measurements carried out for different post-growth annealing conditions and spacer layer thickness suggest that the positive magnetoresistance peaks originate in a noncollinear spin valve effect due to spin-dependent scattering that is believed to occur primarily at interfaces.

\end{abstract}
\maketitle

\section{Introduction}
Heterostructures derived from ferromagnetic semiconductors (FMS) are of current interest for both fundamental and applied problems in semiconductor-based spintronics \cite{Wolf:2001ev}. In analogy with tunnel-based metallic spintronic devices, prior work has studied spin-dependent transport in various tunneling geometries \cite{Chiba:2000apl,Tanaka:2001bc,Chun:2002rx,Chiba:2004ty,Ruster:2005prl} that incorporate the ``canonical'' FMS \gma.\cite{Macdonald:2005rx} Some experiments suggest contributions of spin-dependent scattering (SDS) to spin transport in tunneling FMS devices.\cite{Chiba:2000apl} However, there are no reports of SDS in the {\it conventional} spin valve geometry wherein the FMS layers are separated by a {\it conducting} non-magnetic spacer (for instance, two \gma~ layers separated by non-magnetic p-doped GaAs). In such a device, the observation of a magnetoresistance (MR) that depends on the relative orientation of the magnetizations of the FMS layers would provide an all-semiconductor analog of a phenomenon central to metal-based spintronics. In this paper, we report the observation of such a spin valve effect in \gma/GaAs:Be/\gma~  trilayers. In strong contrast with metallic spin valves, the associated MR in these FMS spin valves is large on an absolute scale ($\Delta R \sim 15 \Omega$), but small on a relative scale ($\frac{\Delta R}{R} \sim 0.2 \%$). In addition, the unusual in-plane anisotropy of \gma~ results in spin valve devices that show MR when the relative magnetization of the two FMS layers switches from collinear to non-collinear (orthogonal) configurations.    

In metallic systems, the conventional spin valve effect yields a giant magnetoresistance (GMR) when the magnetizations of  the ferromagnetic layers switch from parallel to antiparallel states.\cite{Baibich:1988nv,Tsymbal:2001lr} The physical origins of GMR are attributed to differences in the contributions of SDS to the electrical conductivity for majority and minority spin carriers. A detailed knowledge of the band structure, interfaces and impurity scattering is needed to quantitatively understand the GMR effect in a spin valve in the current-in-plane (CIP) geometry. However, a qualitative understanding can be obtained using a simple two resistor model wherein majority and minority spins are assigned different resistivities ($\rho_{\uparrow}$ and $\rho_{\downarrow}$, respectively), and where the spin diffusion length is larger than the spacer layer thickness ($d$). Such an analysis leads to a MR given by: \cite{Tsymbal:2001lr}
\begin{equation}\label{eq1}
\frac{{\Delta R}}{R} = \frac{{(\alpha  - 1)^2 }}{{4(\alpha  + pd_{NM} /d_{FM} )(1 + pd_{NM} /d_{FM} )}}.
\end{equation}
Here, $\alpha = \rho_{\downarrow} /\rho_{\uparrow}$, $p = \rho_{NM}/\rho_{\uparrow}$ and $\rho_{NM}$ is the spacer layer resistivity.  Therefore, in order to obtain GMR, for a given value of $\alpha$, it is important to have a low resistivity of the non-magnetic spacer layer, at least comparable to the majority-spin resistivity $\rho_{\uparrow}$ in the magnetic layers. For the CIP geometry, contributions to the spin valve effect from SDS at interfaces are enhanced by increasing both the mean free path 
($l_e$)  and the spin diffusion length ($l_{sf}$) of carriers in the spacer layer. Both these quantities are short in the low temperature grown p-GaAs spacer incorporated within \gma-based trilayers. We estimate typical Drude values of $l_e \sim 3 - 5$ nm (due to point defects created during the low temperature growth) and $l_{sf} \sim 5 $ nm (due to the short spin scattering time for holes). The spin valve effect can also have some contributions from SDS in the bulk of the ferromagnetic layers. However, both majority and minority holes presumably have a short spin diffusion length in \gma, making scattering in the bulk largely spin-independent and reducing this source of GMR. All these factors suggest a very narrow window of device parameters that would allow the definitive observation of the spin valve effect in \gma-based trilayer devices. Finally, we note that other sources of pronounced MR in \gma~ -- such as the anisotropic magnetoresistance (AMR), the planar Hall effect (PHE) and the anomalous Hall effect\cite{Goennenwein:2005ak,Tang:2003kj,Xiang:2005hp} -- may also complicate the identification of the spin valve effect.  However, as shown later in this paper, the contributions of each of these effects can be distinguished from GMR. Additional MR effects have also been observed in tunneling geometries\cite{Ruster:2005prl} but are unimportant in the conventional spin valve geometry studied here. 

\section {Experimental Details}
We fabricated \gma/GaAs:Be/\gma~ trilayers by low temperature molecular beam epitaxy on semi-insulating (001) GaAs substrates after the growth of a 100-nm-thick standard (high temperature) GaAs buffer layer. The thickness of the \gma~ layers is $\sim 10 \rm{nm}$ (calibrated using reflection high energy electron diffraction techniques) and the Mn concentration is nominally $x \sim 0.03$ (based upon earlier calibrations). The GaAs:Be spacer layer is p-doped using a Be source, with a nominal hole density $p \sim 10^{20} \rm{cm}^{-3}$ and a resistivity of $\sim 10 \rm{m} \Omega ~ \rm{cm}$; this is based upon electrical measurements of thick Be-doped GaAs epilayers grown under identical conditions.  Electrical transport measurements were carried out using dc techniques on lithographically-defined, wet-etched $500 \mu \rm{m} \times 200 \mu \rm{m}$ Hall bar mesas oriented along $[110]$ and $[100]$, with In contacts. We also measured the magnetization ($M$) of each sample using a Quantum Design superconducting quantum interference device (SQUID) magnetometer.

In order to observe spin valve behavior in such trilayers, the \gma~ layers need to have distinct coercive fields (\hc). We accomplish this by post-growth annealing which drives hole-compensating Mn interstitials from the topmost \gma~ layer to benign regions at the free surface of the sample;\cite{Edmonds:2004tz}  in contrast, the diffusion of interstitials in the buried \gma~  layer is believed to be self-limited by the establishment of Coulomb barriers at the GaAs/\gma~ interfaces. This results in clear changes to the magnetic properties of the top \gma~ layer, while leaving the magnetic properties of the buried \gma~ layer unaffected.\cite{Chiba:2003ty,Stone:2003rx}
We investigated samples with a GaAs spacer layer thickness of 2 nm, 5 nm and 10 nm: devices fabricated from these samples are identified as A, B, and C, respectively. We also studied pieces of sample B as a function of annealing time at $190^{\circ}$ C: devices fabricated from pieces of sample B are labeled as  B1, B2, B3, B4, and B5, corresponding to as-grown, and annealed for 30 min, 60 min, 90 min, 120 min, respectively. All devices fabricated from samples A and C are annealed for 60 min. Finally, as a control sample, we measured an as-grown 10 nm thick epilayer of \gma~ with nominally identical Mn concentration as in the trilayers (device D).

Figure 1(a) shows the temperature dependence of the remanent magnetization for devices B1 and B4 (as-grown and 90 minute anneal, respectively). The data indicate that annealing is necessary to create distinct Curie temperature (\tc) for each of the \gma~ layers (\tc = 65 K for the bottom \gma~ layer and \tc = 90 K for the top layer).  Figure 1(b) shows that the annealing also yields a two-step $M(H)$ hysteresis loop (measured with $\vec{H} || [110]$), with a distinct value of \hc~ for each of the \gma~ layers.  Note that the switching of the magnetization of the \gma~ layers occurs with a $90^{\circ} $ rotation between in-plane easy axes ($[ \bar{1}00] \rightarrow [010] \rightarrow [100] \rightarrow [0\bar{1}0]$). \cite{Goennenwein:2005ak,Tang:2003kj} Since the two \gma~ layers do not switch simultaneously,  this results in a plateau in $M(H)$. Temperature- and field-dependent SQUID measurements of devices B1 through B5 show that the magnetic behavior of the top and bottom \gma~ layers gradually becomes more different as a function of annealing time (data not shown).

In order to track the magnetization configurations in the trilayer, we exploit the GPHE in \gma~ by carrying out transport measurements in the current-in-plane (CIP) geometry with the magnetic field $\vec{H}$ deliberately misaligned with respect to the current density $\vec{j} = j \hat{x}$ at an  angle $\vartheta$.  Figures 1(c) - (f) show \rxx$(H)$ and \rxy$(H)$ for devices B1, B2, B3 and B5, respectively at $T = 4.2$ K with $\vec{j} || \hat{x} || [110]$ and $\vartheta \sim 20^{\circ}$.  The data for the as-grown device (B1) are similar to those for a single \gma~ epilayer, because the two \gma~ layers have the same \tc~ and \hc. However, we observe a clear qualitative change in the MR as the device is annealed: with increasing annealing time, \rxx~ shows a tendency to stand out of the MR background, eventually resulting in two pairs of double MR peaks, which must have a different origin from the AMR associated with a single \gma~ epilayer. The annealing-induced changes to \rxx$(H)$ are accompanied by corresponding changes in  \rxy$(H)$, which evolves from a one step hysteresis loop to a two-step hysteresis loop with increasing annealing time.  We will argue below that the double peak MR structure is an unambiguous indication of a spin valve effect that occurs when the magnetizations of the two \gma~ layers are orthogonal to each other. Additional evidence for the spin valve origin of these MR peaks arises from the temperature dependence of \rxx~ and \rxy: in Fig. 2, we show the magnetic field-dependence of \rxx~ and \rxy~ at different temperatures in device B4 with $\vec{j} || \hat{x} || [110]$ and $\vartheta \sim +15^{\circ}$. As the temperature increases, the saturated magnetization and the coercive fields of the two \gma~ layers decrease: as a consequence, there is a decrease in the magnitudes of the jumps in both \rxy$(H)$~ and \rxx$(H)$.  Once the bottom layer becomes paramagnetic ($T >  60 $~K), the spin valve effect vanishes, but the GPHE due to the top \gma~ layer still persists.  

\section {Theoretical Analysis}
We now develop an analytical model of the observed behavior of \rxx$(H)$ and \rxy$(H)$ in the annealed devices. The discussion will refer to the data shown in Fig. 3 for device B4. We start with standard equations that describe the electrical field $\vec{E}$ within a single domain ferromagnet with in-plane magnetization:\cite{Tang:2003kj}
\begin{eqnarray}\label{eq2}
 E_x  = j\rho _ \bot   + j\left( {\rho _{||}  - \rho _ \bot  } \right)\cos ^2 \varphi  \\ 
 E_y  = j\left( {\rho _{||}  - \rho _ \bot  } \right)\sin \varphi \cos \varphi
\end{eqnarray}
Here, $\varphi$ is the angle between the direction of the current density $\vec{j}$ ([110]) and the magnetization $\vec{M}$; $\rho _ \bot$ and $\rho _{||}$ are the resistivities for current flow perpendicular and parallel to the magnetization, respectively. We note that $\rho_{||}$~ is usually smaller than $\rho_\bot$~ in \gma.\cite{Tang:2003kj} We now use these equations to develop an analytical understanding of \rxy$(H)$ in a trilayer device by treating it as three parallel resistors with resistivity $\rho_1$ (top FM layer), $\rho _{NM}$ (non-magnetic spacer layer) and $\rho_2$ (bottom FM layer). 

For an in-plane magnetic field, there is no transverse electric field ($E_y$) due to the ordinary Hall effect. However, there is a contribution to $E_y$ from the PHE in the two FM layers, driving a transverse current in each FM layer. According to eq. 2, if the two layers were independent, such a transverse current would experience a resistivity $\rho _ \bot   + \left( {\rho _{||}  - \rho _ \bot  } \right)\cos ^2 (90^0-\varphi) = \rho _ \bot   + \left( {\rho _{||}  - \rho _ \bot  } \right)\sin ^2 \varphi$ in each layer. Hence, using eq. 3, the PHE in the two FM layers would lead to a net transverse current given by:
\begin{equation}\label{eq3}
j_M =	-\left[ {\sum\limits_{i = 1,2} {\frac{{j_i \left( {\rho _{i||}  - \rho _{i \bot } } \right)\sin \varphi _i \cos \varphi _i }}{{\rho _{i \bot }  + \left( {\rho _{i||}  - \rho _{i \bot } } \right)\sin ^2 \varphi _i }}} } \right].
\end{equation}
Here, $j_1$and $j_2$ are the current densities (along the $x$-direction by definition) in the top and bottom \gma~ layers, respectively. But, the three parallel resistors in the trilayer must experience the same effective transverse field $\left( {E_y} \right)_{\rm{eff}}$ that drives a transverse electrical current $j_E$. Taking each layer's thickness into account, 
\begin{equation}
j_E =\left( {E_y} \right)_{\rm{eff}}\left[ {\frac{1}{{2\rho _{NM} }} + \sum\limits_{i = 1,2} {\frac{1}{{\left( {\rho _{i \bot }  + \left( {\rho _{i||}  - \rho _{i \bot } } \right)\sin ^2 \varphi _i } \right)}}} } \right] 
\end{equation}
A meaningful physical measurement of the Hall effect requires carrier accumulation on the sidewalls at equilibrium (i.e. the total transverse electrical current should be zero for the entire device). Under this condition, $j_E = j_M$ and hence the effective transverse field $(E_y)_{\rm{eff}}$ is given by:
\begin{equation}\label{eq3}
\left( {E_y} \right)_{\rm{eff}}  = \left[ {\sum\limits_{i = 1,2} {\frac{{j_i \left( {\rho _{i||}  - \rho _{i \bot } } \right)\sin \varphi _i \cos \varphi _i }}{{\rho _{i \bot }  + \left( {\rho _{i||}  - \rho _{i \bot } } \right)\sin ^2 \varphi _i }}} } \right]\left[ {\frac{1}{{2\rho _{NM} }} + \sum\limits_{i = 1,2} {\frac{1}{{\left( {\rho _{i \bot }  + \left( {\rho _{i||}  - \rho _{i \bot } } \right)\sin ^2 \varphi _i } \right)}}} } \right]^{ - 1} 
\end{equation}
From Eq. 6, it follows that $(E_y)_{\rm{eff}}$ has a minimum when the two \gma~ layers have their magnetization parallel ($\varphi_1 = \varphi_2 = 45^{\circ}$). In the down-field sweep in Fig. 3(b), this corresponds to the field range between ``A" and $H \sim -0.05$ kOe. In contrast, $(E_y)_{\rm{eff}} \sim 0$ when the magnetization of the top layer is switched by the external magnetic field ($\varphi_1 = 135^{\circ}$ and ($\varphi_2 = 45^{\circ}$), corresponding to ``B" in the down sweep in Fig. 3(b). Finally, $(E_y)_{\rm{eff}}$ reaches a maximum when the bottom layerÕs magnetization switches ($\varphi_1 = \varphi_2 = 135^{\circ}$), corresponding to ``C" in the down sweep in Fig. 3(b).  At ``D," we again have $(E_y)_{\rm{eff}} \sim 0$ with the magnetizations orthogonal to each other as in ``B." The abrupt switching of the magnetization of each layer in the trilayer system results in a characteristic change of \rxy~ that allows us to track the relative configuration of magnetization orientations. As shown in Fig. 3, the distinctive features in \rxx$(H)$ are then directly correlated with the magnetization configuration of the \gma~ layers, with a minimum \rxx~ when the magnetizations are aligned parallel, and an enhanced \rxx~ for non-collinear configurations.\cite{LI:1995ng,SLONCZEWSKI:1991ne} 

However, Fig. 3 also shows that the detailed shape of the MR in these FMS spin valves is more complex than that found in most metallic spin valves, showing a double peak structure. This feature arises from an interesting interplay between the spin valve effect and the intrinsic AMR of an individual \gma~ layer, leading to a MR {\it valley} exactly where a pure spin valve effect would lead to a MR peak. This behavior may be qualitatively explained using the magnetoimpurity scattering model \cite{Goennenwein:2005ak,Nagaev:2001ue} which shows that \rxx$(H)$ for a single \gma~ layer decreases abruptly when the magnetization first switches upon field reversal (as in Fig. 1 (c)). This occurs because of a discontinuous change in $\rho _ \bot   = a - b |\vec{B}| = a - b | \vec{H} + \vec{M} |$, where both $a$ and $b$ are positive parameters. Hence, when the magnetization of the top \gma~ layer in a spin valve device switches, $|\vec{B}| = |\vec{H}+ \vec{M}| $ increases, $\rho _ \bot$ decreases and \rxx~ decreases (Eq. 1). As a consequence, the enhanced MR due to the spin valve effect is opposed by the intrinsic single layer AMR precisely when the layers switch from parallel to orthogonal configurations, resulting in a MR valley. 

Yet more evidence for the spin valve effect arises from studies of MR along a different crystalline direction ([010]). In this configuration, for each individual layer of \gma, the angle between the current density and magnetization ($\varphi$ in Eq. 3) is restricted to $0^{\circ}$, $90^{\circ}$, $180^{\circ}$ and $270^{\circ}$ degrees because the magnetic easy axis has to lie along one of the $\langle 100 \rangle$ directions. Thus, eq. 3 predicts that the abrupt switching of the magnetization in a single layer will be accompanied by an equally abrupt change of longitudinal resistance $\delta R_{x} \propto (\rho_{\bot} - \rho_{||})$.  In a trilayer device, such switching produces four distinct jumps in longitudinal resistance as the magnetic field is swept from one field orientation to the opposite one. This behavior is seen in Fig. 4 which shows \rxx$(H)$ at $T = 1.5$ K in device B4 with $\vec{j} || [010]$ and $H$ misaligned at $\sim +15^{\circ}$ with the $[110$] direction. We focus on the jumps labeled as ``I" and ``II" in the Fig. 4. These correspond respectively to the switching of the magnetization of the top layer from a direction antiparallel to the field to orthogonal to the field and from orthogonal to parallel to the field. We model the longitudinal resistance of the trilayer device as three parallel resistors: $R^{-1} = \Sigma (R_{i})^{-1}$.  Since the observed $\delta R/R$ in Fig. 4 is relatively small (less than $1.7 \%$), the total resistance of the trilayer system and the single layer resistances can be approximately treated as constants. Hence, if the only contribution to MR at jumps 1 and II is due to the switching of the top FMS layer, then $\Delta R = \frac{R^2}{{R_{\rm{top}}}^2} \Delta R_{\rm{top}}$.  If there are no contributions from SDS and one only considers the change in $\rho_\bot$ due to magnetization switching, then the two jumps in resistance due to the switching of the top layer's magnetization (marked as I and II in Fig. 4) should have a distinct relationship: specifically, jump I should be a little smaller than jump II. This is because at jump I, $\rho_\bot$ opposes the AMR contribution, but at jump II, it enhances the AMR contribution. However, Fig. 4 shows exactly the opposite behavior:  the resistance change is bigger at jump I than at jump II. Further, the difference in magnitude of the resistance jumps ($\sim 12 \Omega$) cannot be explained by AMR alone. On the other hand, if the spin valve effect is taken into account, it will enhance the positive resistance jump but oppose the negative resistance jumps. Hence, the presence of a spin valve effect is consistent with our observation that  jump I is bigger than jump II. 

\section {Discussion}
The spin valve effect in these semiconductor-based trilayers could have two possible microscopic origins: SDS occurring at the interfaces between layers and SDS in the bulk of the FMS layer and possibly due to diffused magnetic ions in the non-magnetic spacer layer. From the data presented here, we cannot definitively identify the source of SDS, although it is reasonable to expect that SDS in the bulk of the FMS layers does not play a dominant role given the short the diffusion length for holes is in \gma. The observed enhancement of the spin valve effect by annealing is consistent with SDS at the interfaces: we speculate that the annealing leads to compositional disorder at the interfaces (possibly due to migration of both Mn and Be atoms), resulting in enhanced interfacial SDS. Such a mechanism has been suggested as an explanation for annealing-induced increases to GMR in metallic multilayers, although we caution that compositional disorder may also decrease GMR under certain conditions.\cite{Jonkers:2001op}  Within such a picture, the spin valve effect should show a non-monotonic dependence on annealing time due to the trade-off between the enhancement of SDS due to Mn diffusion into the spacer layer and reduction of spin scattering due to Be migration in to the \gma~ layers. This is supported by our data, showing that 90-minute annealing has a bigger spin valve effect than shorter and longer annealing times. 

The role of SDS at interfaces is also consistent with the dependence of the observed spin valve effect on the thickness of the spacer layer. This is shown in Figs. 5 (a) - (d) where we compare the field dependence of \rxx~ for devices A, B3, C (all annealed for 60 minutes at $190 ^{\circ}$ C) and the (as-grown) control sample D. The data show that the spin valve effect is absent in devices with the thinnest (2 nm) spacer layer (Fig. 5(a)), where the behavior is similar to that of the control epilayer (Fig. 5(d)).  We attribute this to a coupling between the \gma~ layers which makes the trilayer act like a single layer. The spin valve effect is strongest in devices with the 5 nm spacer (Fig. 5(b)) and absent in devices with the thickest (10 nm) spacer (Fig. 5(c)). The latter is consistent with an anticipated hole spin diffusion length of several nm. We note that qualitative features of this spacer thickness dependence has been reproduced in devices fabricated from different wafer runs of nominally identical samples.

\section {Conclusion}
In conclusion, we have shown that annealing is critical to the observation of GMR in all-semiconductor \gma/Be:GaAs/\gma~ trilayer devices. The dependence of the spin valve effect on annealing time and spacer thickness suggests that the spin valve effect principally arises from interfacial spin-dependent scattering. These semiconductor spin valves give access to a different regime of parameter space for studying GMR compared to metallic systems and may provide fundamental new insights into the underlying physics. Further insights into the spin valve effect in these all-semiconductor trilayer devices can be obtained in the current-perpendicular-to-the-plane geometry and such measurements will be reported in future reports. 
This research has been supported by grant numbers ONR N0014-05-1-0107, University of California-Santa Barbara subcontract KK4131, and NSF DMR-0305238 and -0401486. We also acknowledge the use of the NSF funded NNUN facility at Penn State University.

\newpage
\begin{center}
{\bf References}
\end{center}

\newpage

\begin{center}
{\bf Figure Captions}
\end{center}

FIG. 1. Magnetization of sample B as-grown and after annealing at $190^{\circ}$C for 90 min as a function of (a) temperature and (b) magnetic field.  Remanent magnetization in (a) was measured in a field of 30 Oe after cooling in a 10 kOe field. Inset in (b) shows magnetic field and 
magnetization configuration at the flat step in $M(H)$ hysteresis loop. 
Magnetic field dependence of \rxx~ and \rxy~ at $T = 4.2$ K  in devices fabricated from the same wafer, but subject to different annealing periods is shown in (c) Device B1: as grown; (d) device B2: annealed for 30 min; (e) device B3: annealed for 60 min; (f) device B5: annealed for 120 min. The data are all taken in the CIP geometry with $\vec{j} || [110]$ and $\vec{H}$ aligned $20^{\circ}$ off [110].

FIG. 2. Temperature dependence of of (a) \rxx~ and (b) \rxy~ in device B4.The data are taken in the CIP geometry with $\vec{j} || [110]$ and $\vec{H}$ aligned $15^{\circ}$ off [110]. The data are offset vertically for clarity.

FIG. 3. Magnetic field dependence of (a) \rxx~ and (b) \rxy~ at $T = 1.5$~K in device B4. The data are taken in the CIP geometry with $\vec{j} || [110]$ and $\vec{H}$ aligned $15^{\circ}$ off [110]. Arrows in (b) represent relative magnetizations of the top and bottom \gma~ layers. Panels (c) and (d) depict the configurations of magnetization and magnetic field when  $\varphi = 45^{\circ}$ and  $\varphi = 135^{\circ}$, respectively. The sample was annealed for 90 minutes at $190^{\circ}$C.

FIG. 4. Magnetic field dependence of \rxx~ in sample B3 at 1.5 K. The data are taken in CIP geometry with $\vec{j} ||[010]$ and H aligned $15^{\circ}$ off [110]. The sample was annealed for 90 minutes at $190^{\circ}$C.

FIG. 5. Magnetic field dependence of \rxx~ and \rxy~ at $T = 4.2$~K  in trilayer devices with (a) 2 nm spacer (device A), (b) 5 nm spacer (device B3) and (c) 10 nm spacer (device C). All three trilayers are subject to 60 minute annealing at $190^{\circ}$C. Panel (d) shows data for the as-grown epilayer (device D). The data are all taken in the CIP geometry with $\vec{j} || [110]$ and $\vec{H}$ aligned $20^{\circ}$ off [110].

\newpage

\begin{figure}[h]
\begin{center}
\includegraphics[]{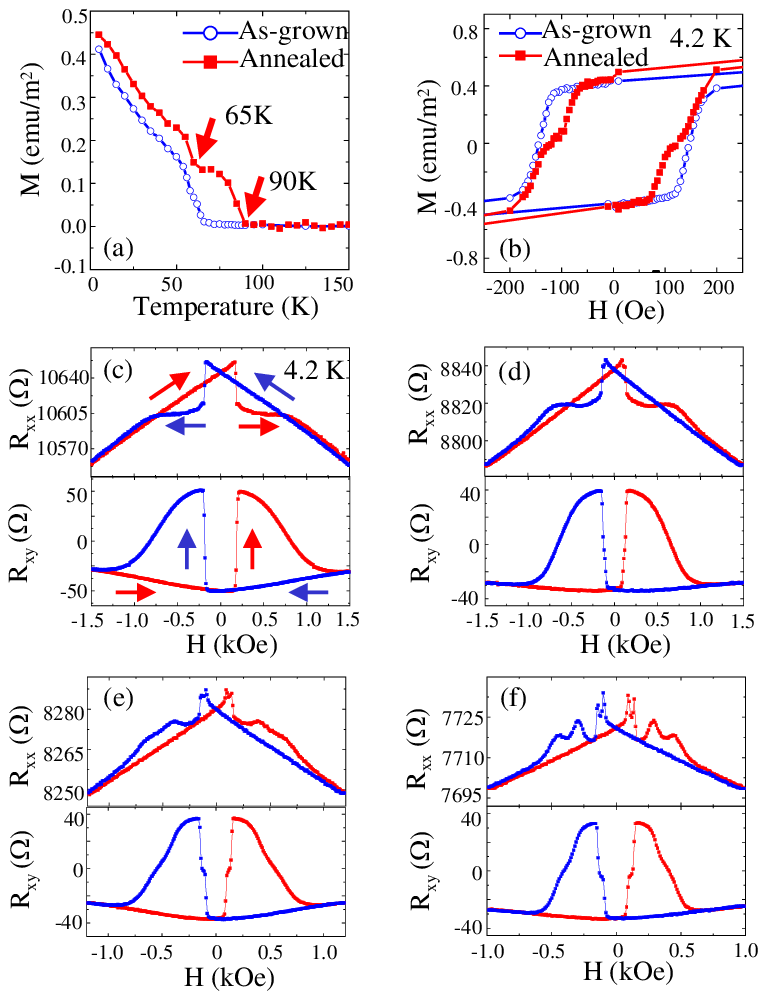}
\label{xiang_fig1}
\end{center}
\caption{Xiang et al.}
\end{figure}

\newpage
\begin{figure}[h]
\begin{center}
\includegraphics[]{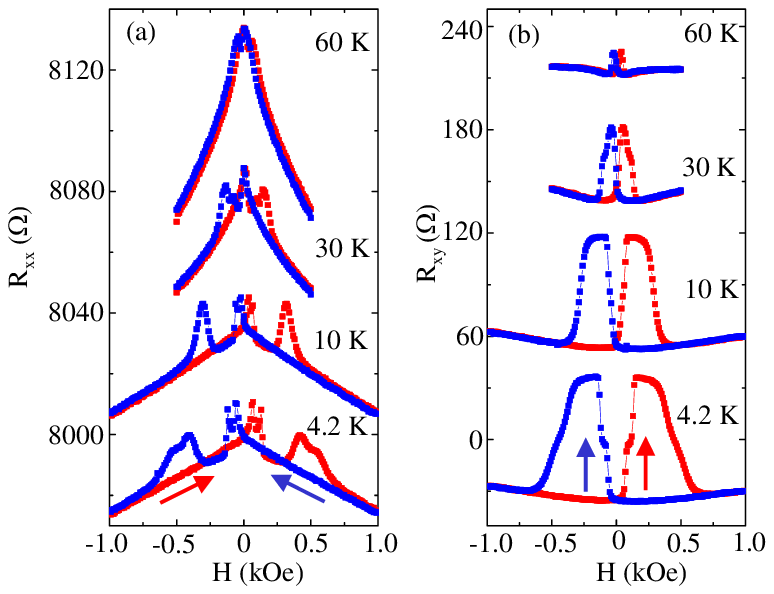}
\label{xiang_fig2}
\end{center}
\caption{Xiang et al.}
\end{figure}

\newpage
\begin{figure}[h]
\begin{center}
\includegraphics[]{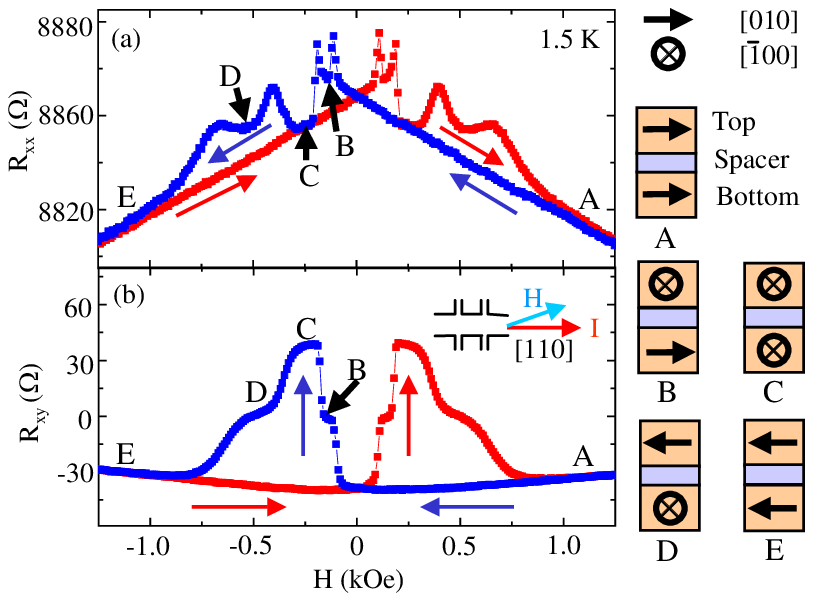}
\label{xiang_fig3}
\end{center}
\caption{Xiang et al.}
\end{figure}

\newpage
\begin{figure}[h]
\begin{center}
\includegraphics[]{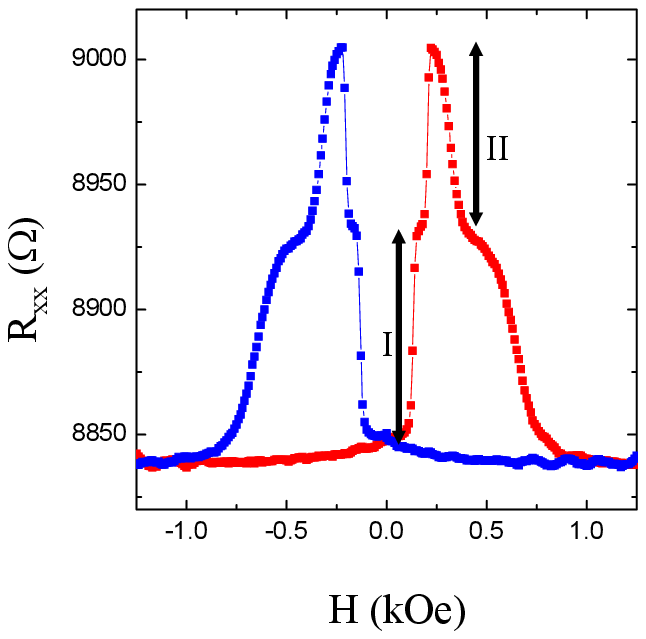}
\label{xiang_fig5}
\end{center}
\caption{Xiang et al.}
\end{figure}

\newpage
\begin{figure}[h]
\begin{center}
\includegraphics[]{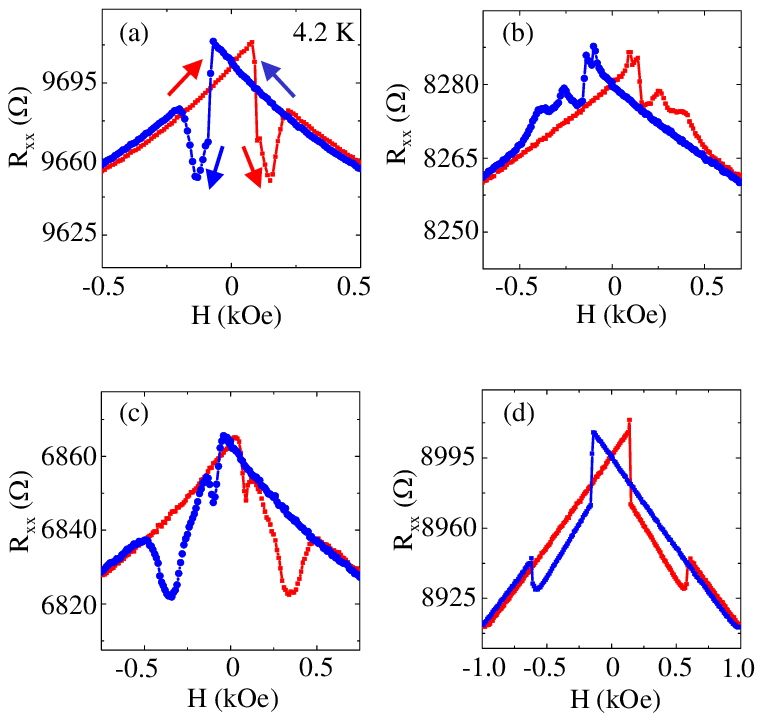}
\label{xiang_fig4}
\end{center}
\caption{Xiang et al.}
\end{figure}

\end{document}